\documentclass[12pt,epsf]{article}
\usepackage{graphicx}
\begin{document}

\def\a{\alpha}
\def\b{\beta}
\def\c{\varepsilon}
\def\d{\delta}
\def\e{\epsilon}
\def\f{\phi}
\def\g{\gamma}
\def\h{\theta}
\def\k{\kappa}
\def\l{\lambda}
\def\m{\mu}
\def\n{\nu}
\def\p{\psi}
\def\q{\partial}
\def\r{\rho}
\def\s{\sigma}
\def\t{\tau}
\def\u{\upsilon}
\def\v{\varphi}
\def\w{\omega}
\def\x{\xi}
\def\y{\eta}
\def\z{\zeta}
\def\D{\Delta}
\def\G{\Gamma}
\def\H{\Theta}
\def\L{\Lambda}
\def\F{\Phi}
\def\P{\Psi}
\def\S{\Sigma}

\def\o{\over}
\def\beq{\begin{eqnarray}}
\def\eeq{\end{eqnarray}}
\newcommand{\gsim}{ \mathop{}_{\textstyle \sim}^{\textstyle >} }
\newcommand{\lsim}{ \mathop{}_{\textstyle \sim}^{\textstyle <} }
\newcommand{\vev}[1]{ \left\langle {#1} \right\rangle }
\newcommand{\bra}[1]{ \langle {#1} | }
\newcommand{\ket}[1]{ | {#1} \rangle }
\newcommand{\EV}{ {\rm eV} }
\newcommand{\KEV}{ {\rm keV} }
\newcommand{\MEV}{ {\rm MeV} }
\newcommand{\GEV}{ {\rm GeV} }
\newcommand{\TEV}{ {\rm TeV} }
\def\diag{\mathop{\rm diag}\nolimits}
\def\Spin{\mathop{\rm Spin}}
\def\SO{\mathop{\rm SO}}
\def\O{\mathop{\rm O}}
\def\SU{\mathop{\rm SU}}
\def\U{\mathop{\rm U}}
\def\Sp{\mathop{\rm Sp}}
\def\SL{\mathop{\rm SL}}
\def\tr{\mathop{\rm tr}}

\def\IJMP{Int.~J.~Mod.~Phys. }
\def\MPL{Mod.~Phys.~Lett. }
\def\NP{Nucl.~Phys. }
\def\PL{Phys.~Lett. }
\def\PR{Phys.~Rev. }
\def\PRL{Phys.~Rev.~Lett. }
\def\PTP{Prog.~Theor.~Phys. }
\def\ZP{Z.~Phys. }

\def\sla#1{\rlap/#1}

\baselineskip 0.7cm

\begin{titlepage}

\begin{flushright}
ICRR-Report-613-2012-2\\
IPMU-12-0035\\
UT-12-08\\
\end{flushright}

\vskip 1.35cm
\begin{center}
{\large \bf
A 125\,GeV Higgs Boson Mass and Gravitino Dark Matter
\\in R-invariant Direct Gauge Mediation
}
\vskip 1.2cm
Masahiro Ibe$^{1,2}$ and Ryosuke Sato$^{2,3}$
\vskip 0.4cm

{\it
$^1$ICRR, University of Tokyo, Kashiwa 277-8582, Japan\\
$^2$Kavli IPMU, University of Tokyo, Kashiwa 277-8583, Japan\\  
$^3$Department of Physics, University of Tokyo, Tokyo 113-0033, Japan\\
}

\vskip 1.5cm

\abstract{
We discuss the Standard Model-like Higgs boson mass in the Supersymmetric Standard Model
in an R-invariant direct gauge mediation model with the gravitino mass in the ${\cal O}(1)$\,keV range.
The gravitino dark matter scenario in this mass range 
is a good candidate for a slightly warm dark matter.
We show that the Higgs boson mass around 125\,GeV
suggested by the ATLAS and CMS experiments
can be easily achieved in R-invariant direct gauge mediation
models with the gravitino mass in this range.
}
\end{center}
\end{titlepage}

\setcounter{page}{2}

\section{Introduction}
The light gravitino is one of the most motivated 
candidate of dark matter in the supersymmetric 
theories, since the gravitino is the unique and the inevitable prediction 
of supergravity.
If the gravitino were in the thermal bath in early universe
the observed dark matter density, $\Omega_{\rm DM}h^2 \sim 0.1$, requires
the gravitino mass to be $m_{3/2} \simeq 100\,{\rm eV}$.
Unfortunately, however, such a very light gravitino decouples from the thermal bath 
when it is still relativistic, and the resultant free-streaming length is too long
to be consistent with the successful galaxy formation\,\cite{Viel:2005qj}.

If we had late time entropy production, 
on the other hand,
the thermally produced gravitino dark matter scenario can be viable.
That is, with  late time entropy production, the dark matter density is diluted, and hence,
the relic density of  heavier gravitino dark matter can be consistent 
with the observed dark matter density.
In particular, the gravitino dark matter with a mass in the keV range
which has a free-streaming wavenumber around $k_{FS} \simeq 100-300$\,Mpc$^{-1}h$ 
is drawing  attention as a solution to the seeming discrepancies 
between the observation and the simulated results of the galaxy formation 
based on the cold dark matter scenario\,\cite{WDM}.

In a collaboration with Yanagida and Yonekura\,\cite{Ibe:2010ym}, 
the authors showed that the required entropy production can be 
provided by decays of long lived particles in a supersymmetry breaking sector
based on the R-invariant direct gauge mediation models developed in 
Refs.\,\cite{Izawa:1997gs, Nomura:1997uu}.%
\footnote{See Refs.\,\cite{Fujii:2003iw, Hasenkamp:2010if} for other mechanisms for entropy production
after the decoupling of gravitino. }
Interestingly, 
the direct gauge mediation models tend
to predict a hierarchical sparticle spectrum 
where the scalar fermions are much heavier than the gauginos\,\cite{Izawa:1997gs, Nomura:1997uu}.%
\footnote{
For more recent generic discussions on the R-invariant direct 
mediation models, see Refs.\,\cite{Ibe:2005xc,Shih:2007av,Komargodski:2009jf,Sato:2009dk}.
}
Especially, the model predicts the squark masses in the ${\cal O}(10-100)\,\TEV$ range
for $m_{3/2}={\cal O}(1-10)\,\KEV$. 

As is well studied, such a heavy squark leads to relatively heavy Standard Model  (SM) like Higgs 
boson\,\cite{Okada:1990gg}.
In this Letter, we show that the 
Standard Model-like Higgs boson with a mass around 125 GeV
suggested by the ATLAS \cite{:2012si} and CMS \cite{Chatrchyan:2012tx} experiments
can be easily achieved due to the heavy sfermions predicted
in the R-symmetric direct gauge mediation models 
for $m_{3/2}={\cal O}(1)\,\KEV$, which predicts 
the free-streaming wavenumber around $k_{FS} \simeq 100-300$\,Mpc$^{-1}h$. 

We also discuss how the electroweak symmetry breaking 
is achieved in the R-invariant gauge mediation model 
where the sfermion masses are of $O(10-100)$\,TeV.
As we will show, it is rather difficult to achieve correct 
electroweak symmetry breaking in the minimal supersymmetric standard model
(MSSM) with gauge mediated supersymmetry breaking for squark masses in $O(10-100)$\,TeV.
We discuss possible extension of the supersymmetric standard model
so that correct electroweak symmetry breaking is achieved.

The organization of the Letter is as follows.
In section\,\ref{sec:model}, we discuss the spectrum
of the superparticles in the R-invariant direct gauge mediation for $m_{3/2}=10-100$\,keV.
In section\,\ref{sec:wdm}, we discuss the free-streaming length of  gravitino dark matter.
In section\,\ref{sec:higgsmass}, we discuss the mass of the SM-like Higgs boson mass in the present model.
In section\,\ref{sec:ew}, we discuss how the electroweak symmetry breaking 
is achieved in the R-invariant gauge mediation model for $m_{3/2} = O(1-10)$\,keV.
The final section is devoted to our conclusions.

\section{R-invariant Direct Gauge Mediation Model}\label{sec:model}
\begin{figure}[tbp]
 \begin{minipage}{0.5\hsize}
  \begin{center}
   \includegraphics[width=70mm]{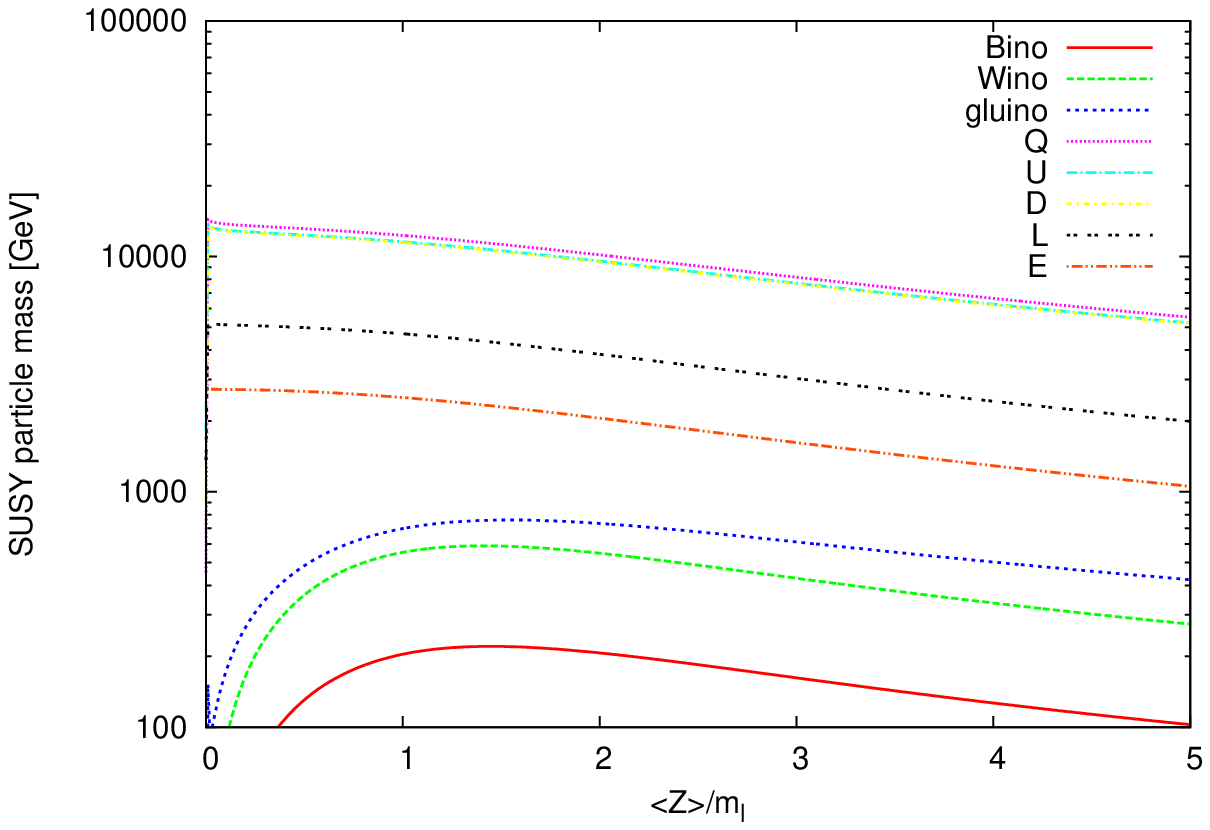}
  \end{center}
 \end{minipage}
 \begin{minipage}{0.5\hsize}
  \begin{center}
   \includegraphics[width=70mm]{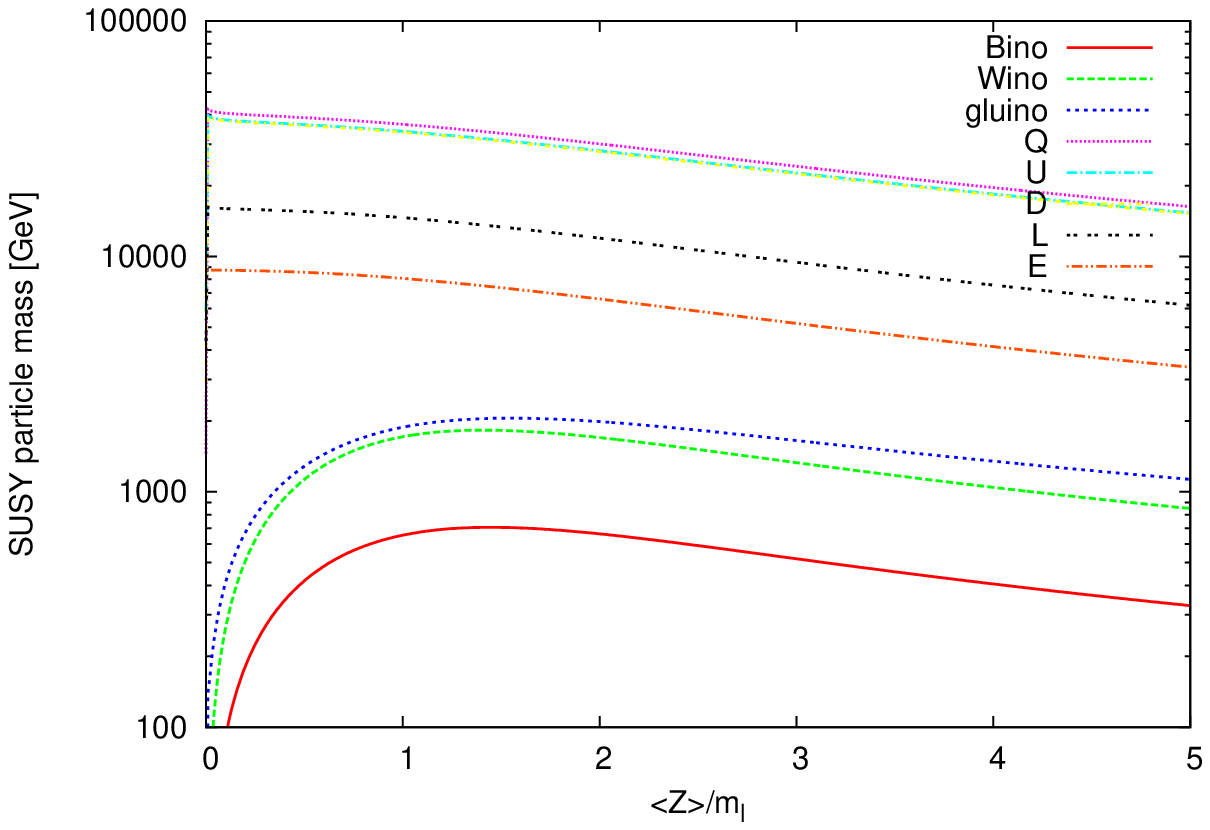}
  \end{center}
 \end{minipage}
  \caption{The superparticle mass in one pair of messengers model with $m_{3/2}$ = 1 keV (left) and $m_{3/2}$ = 10 keV (right) as functions of $\langle Z\rangle / m_\ell$.
  We take $k_\ell F / m_\ell^2 = 0.90$.}
  \label{fig:onepair_mass}
\end{figure}

In this section, we briefly review the simplest R-invariant gauge mediation models developed in 
Refs.\,\cite{Izawa:1997gs, Nomura:1997uu}.
We introduce two pairs of messenger fields $\Psi_i$ and $\tilde \Psi_i$ ($i = 1,2$),
which transform as ${\bf 5}$ and ${\bf 5}^*$ under the $SU(5)$  gauge group
of the grand unified theory (GUT), respectively.
We also introduce a supersymmetry breaking gauge singlet field $Z$
which has non-vanishing expectation values of the $A$ and $F$ terms, such that,
\beq
\langle Z(x,\theta)\rangle = \langle Z\rangle + F\theta^2.
\eeq
In the direct gauge mediation model, we assume this $F$ term is the dominant component 
of the supersymmetry breaking which leads to the gravitino mass,
\beq
m_{3/2} = \frac{F}{\sqrt{3}M_{\rm Pl}},
\eeq
where $M_{\rm Pl} \simeq 2.4\times 10^{18}\,\GEV$ is the reduced Planck mass.
The superpotential of the messenger fields is assumed to be 
\beq
\label{eq:super1}
W = 
\left(
\begin{array}{cc}
\tilde\Psi_1 & \tilde\Psi_2
\end{array}
\right)
\left(
\begin{array}{cc}
kZ & m\\
m & 0
\end{array}
\right)
\left(
\begin{array}{c}
\Psi_1\\
\Psi_2
\end{array}
\right)\ ,
\eeq
where $k$ denotes a coupling constant and $m$ the mass parameter.
We can see the above superpotential is invariant
under the R-symmetry with the charge assignment, $Z(2)$, $\Psi_1(0)$, $\tilde\Psi_1(0)$, $\Psi_2(2)$, $\tilde\Psi_2(2)$.

We split $\Psi_i$ and $\tilde\Psi_i$ into
the SM gauge group $SU(3)_C\times SU(2)_L\times U(1)_Y$ representations,
such as, $\Psi \to (\Psi_d, \Psi_\ell)$ and $\tilde\Psi \to (\tilde\Psi_d, \tilde\Psi_\ell)$.
They transform as $({\bf 3}_{\bf -1/3}, {\bf 2}_{\bf 1/2})$
and $({\bf \bar 3}_{\bf 1/3}, {\bf 2}_{\bf -1/2})$
under the SM gauge groups, respectively.
We denote $\Psi_d$ and $\tilde\Psi_d$ as ``down-type",
$\Psi_\ell$ and $\tilde\Psi_\ell$ as ``lepton-type" messengers.
We also distinguish the coupling constants and the mass parameters 
in Eq.\,(\ref{eq:super1}) for each messenger
by the subscripts such that $k_d$ and $k_\ell$, $m_d$ and $m_\ell$, respectively.
In our analysis, we impose the following relations at the GUT scale:
\beq
k_d = k_\ell\, ,~~~~~
m_d = m_\ell\, ,
\eeq
which are eventually violated at the lower energy scale 
due to the renormalization group evolution.

\begin{figure}[tbp]
 \begin{minipage}{0.5\hsize}
  \begin{center}
   \includegraphics[width=70mm]{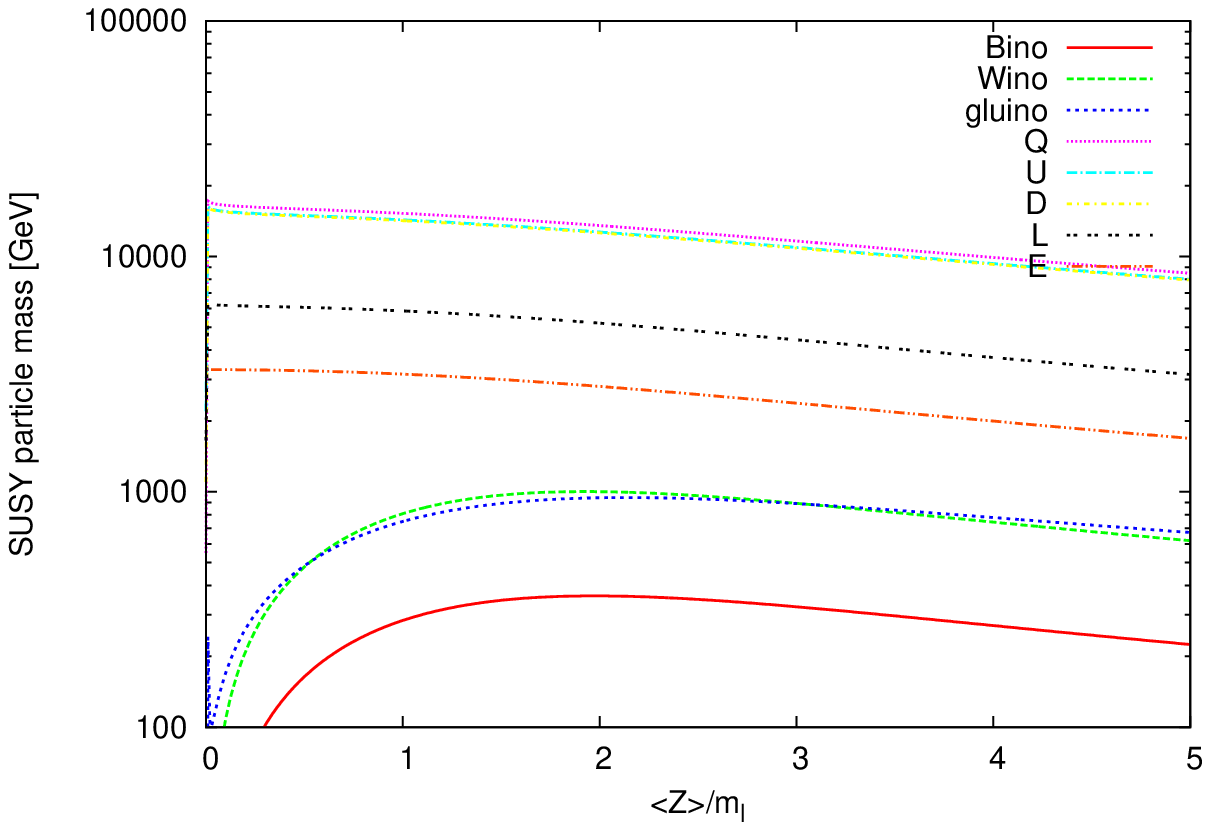}
  \end{center}
 \end{minipage}
 \begin{minipage}{0.5\hsize}
  \begin{center}
   \includegraphics[width=70mm]{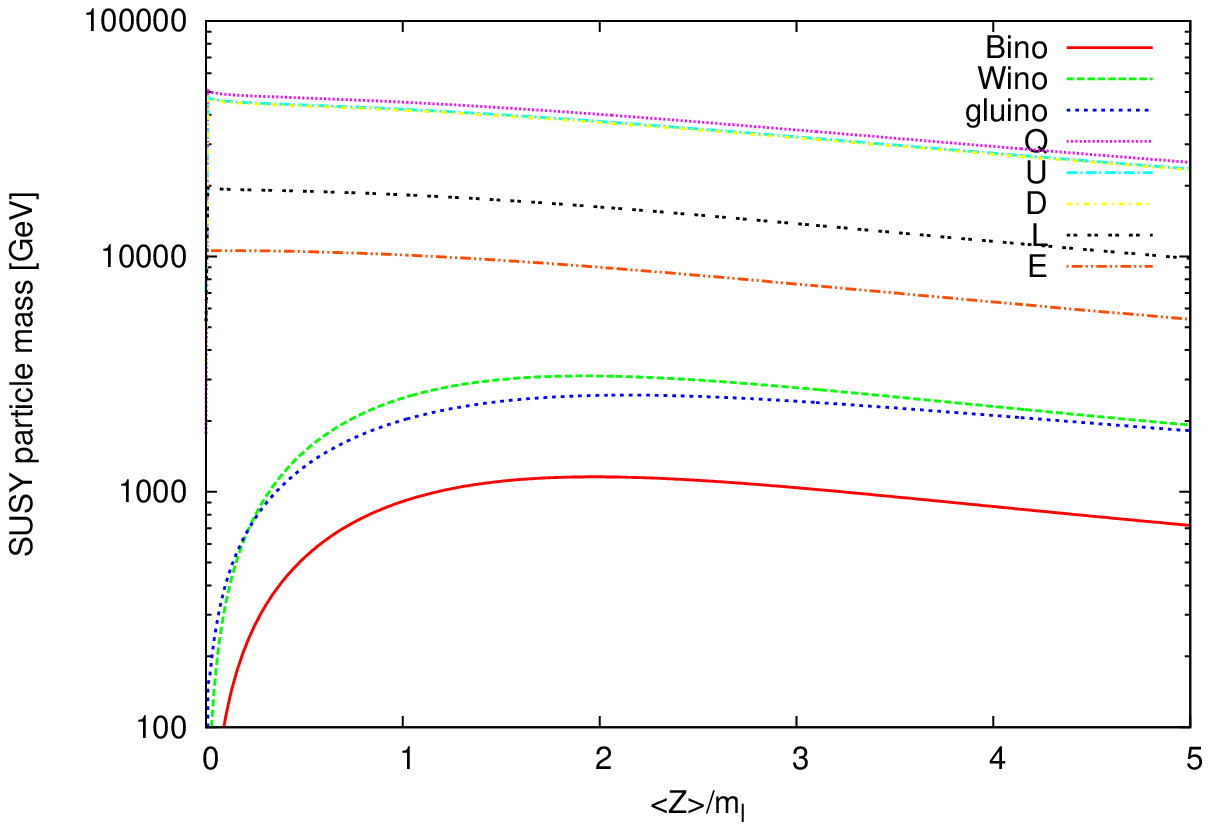}
  \end{center}
 \end{minipage}
  \caption{The superparticle mass in two pairs of messengers model with $m_{3/2}$ = 1 keV and $m_{3/2}$ = 10 keV (right) as functions of $\langle Z\rangle / m_\ell$.
  We take $k_\ell F / m_\ell^2 = 0.90$.}
  \label{fig:twopair_mass}
\end{figure}

The  characteristic features of the mass spectrum are as follows:
\begin{itemize}
\item There is a hierarchy between gaugino masses and sfermion masses.
Because the gaugino masses have no term ${\cal O}(kF/m)$,
the gaugino masses are much suppressed compared to the sfermion masses.
\item Due to the deviation from the GUT  relation $k_d = k_\ell$ and $m_d = m_\ell$ at 
the messenger scale, the gaugino masses do not obey the so-called GUT relation.
Namely, in this model, $M_1 : M_2 : M_3 \neq \a_1 : \a_2 : \a_3 \simeq 1 : 2 : 6$.
\end{itemize}
For detailed discussion on the spectrum, see Refs.\,\cite{Nomura:1997uu, Ibe:2010ym}.
In Fig.\,\ref{fig:onepair_mass}, we show the superparticle mass spectrum in the present model
for $m_{3/2}=1$\,keV and $m_{3/2}=10$\,keV.
In Fig.\,\ref{fig:twopair_mass}, we also show the mass spectrum in the model with 
additional messenger pair.
The figures show that the squark masses are in tens TeV range, while
the gaugino masses are within a TeV range.%
\footnote{
The gluino mass for the model with 
 $m_{3/2}=1$\,keV (the left panel of Fig.\,\ref{fig:onepair_mass})
 is almost excluded by the missing transverse momentum with jets 
 by the ATLAS experiment\,\cite{SUSY}.
}

\section{Slightly Warm Gravitino Dark Matter}\label{sec:wdm}
In this section, we briefly review the property of the gravitino dark matter.
We assume that late time entropy production dilutes the gravitino relic abundance\,\cite{Ibe:2010ym}.
In this case, the resultant gravitino relic density is given by 
\beq
\Omega_{3/2} h^2 \simeq 0.1 \left(\frac{100}{g_*(T_D)}\right) \left(\frac{m_{3/2}}{100~{\rm eV}}\right) 
\D^{-1}\ ,
\label{eq:abundance}
\eeq
Here, $\D$ is the dilution factor due to the late time entropy production,
and $g_*(T_D) \simeq 100$ denotes the effective massless degree of freedom in the thermal bath
at the decoupling temperature $T_D$.
To achieve the observed dark matter density, $\Omega_{DM}h^2 \simeq 0.1$\,\cite{Komatsu:2010fb},
the required dilution factor is given by 
\begin{eqnarray}
\label{eq:dilution}
\D\simeq \left(\frac{100}{g_*(T_D)}\right) \left(\frac{m_{3/2}}{100~{\rm eV}}\right) 
\left(
\frac{0.1}{\Omega_{3/2}h^2}
\right)
\ .
\end{eqnarray}
In Refs.\,\cite{Ibe:2010ym}, we showed that the required dilution factor 
$\D$ can be provided by the decay of long lived heavy particles
in the supersymmetry breaking sector which generates $\vev{S(x,\h)}$ dynamically
based on the dynamical supersymmetry breaking sector developed in Ref.\,\cite{Izawa:1996pk,Intriligator:1996pu}.

The ``warmness'' of a warm dark matter can be measured by the free streaming length 
$\l_{\rm FS}$\,\cite{Bond:1980ha}.
The density perturbation of the scale smaller than $\l_{\rm FS}$ is suppressed.
The free-streaming wavenumber of the gravitino is roughly given by%
\footnote{
Here, we defined  the free-streaming wavenumber\,\cite{Kamada} 
\begin{eqnarray}
k_{FS} = \left(\frac{3 a^2H_0^2 \Omega_{\rm DM}}{2 \vev{v^2}}  \right)^{1/2}\ ,
\end{eqnarray}
at the matter-radiation equality time.
}
\beq
k_{\rm FS} = \frac{2\pi}{\l_{\rm FS}}
\simeq 15\,{\rm Mpc}^{-1} h \left(\frac{m_{3/2}}{1~{\rm keV}}\right)
 \left(\frac{g_*(T_D)}{100} \right)^{1/3}
  \D^{1/3}
\, .
\label{eq:freestreaming}
\eeq
By using Eqs.\,(\ref{eq:dilution}) and (\ref{eq:freestreaming}), we obtain
\beq
k_{\rm FS} \simeq 33\,{\rm Mpc}^{-1} h\left(\frac{m_{3/2}}{1~\KEV}\right)^{4/3}
\left( \frac{\Omega_{3/2} h^2}{0.1} \right)^{-1/3} \ .
\eeq

The warm dark matter with the free-streaming wave number in this range is drawing attention 
as a solution for the small scale crisis
in the conventional cold dark matter ($\L$CDM) model.%
\footnote{
 In the present scenario, the gravitino warm dark matter
 has a Fermi-Dirac momentum distribution rescaled 
 by the dilution factor.
At this time, however, the detailed shape of the momentum distributions 
is not relevant to discuss the small scale 
crisis due to the limited resolutions of 
the $N$-body simulations and the observations.
(See also model independent treatment of the warm dark matter
in Ref.\,\cite{deVega:2009ku}.)
 }
Despite its success in reproducing the large scale structure of the universe, 
several intriguing discrepancies have been reported
between the simulations of structure formation and observations 
at the small scales  in the $\L$CDM model.
The observed number of galactic satellites in the inner region of the halo is an order of magnitude smaller than the predictions of the $\L$CDM model by the 
$N$-body simulations\,\cite{Klypin:1999uc,Strigari:2007ma}.
The $\L$CDM  model also predicts a cuspy dark matter density profile 
in the central region of the halo while the observations 
indicate nearly uniform density cored profile\,\cite{WDM2}.
The warm dark matter with the free-streaming wave number of $k_{FS} = O(10)$\,Mpc$^{-1}h$ is expected 
 to resolve these issues  since it can smear out the density fluctuation in small scale\,\cite{WDM}.
 It should be noted that such a smeared small-scale structure might be probed 
 in future experiments by, for example, using strong gravitational lensing\,\cite{Dalal:2002su},
or by observing the $21$ cm fluctuations in high red-shift region\,\cite{Loeb:2003ya}.

\section{The Lightest Higgs Boson Mass}\label{sec:higgsmass}
In this section, we calculate the SM-like Higgs boson mass in the R-invariant gauge mediation model
for $m_{3/2}=1-100$\,keV where the sfermions are in the tens TeV range.
With such heavy sfermion masses,  the renormalization group improved calculation is required for
the precise calculation of the Higgs mass, with which
we calculate the quartic coupling of the potential of the SM Higgs doublet at the electroweak scale, 
\beq
V = \frac{\l}{2}(|H|^2 - v^2)^2,
\eeq
where $v = \langle H \rangle \simeq 174\,\GEV$.

\begin{figure}[tbp]
  \begin{center}
   \includegraphics{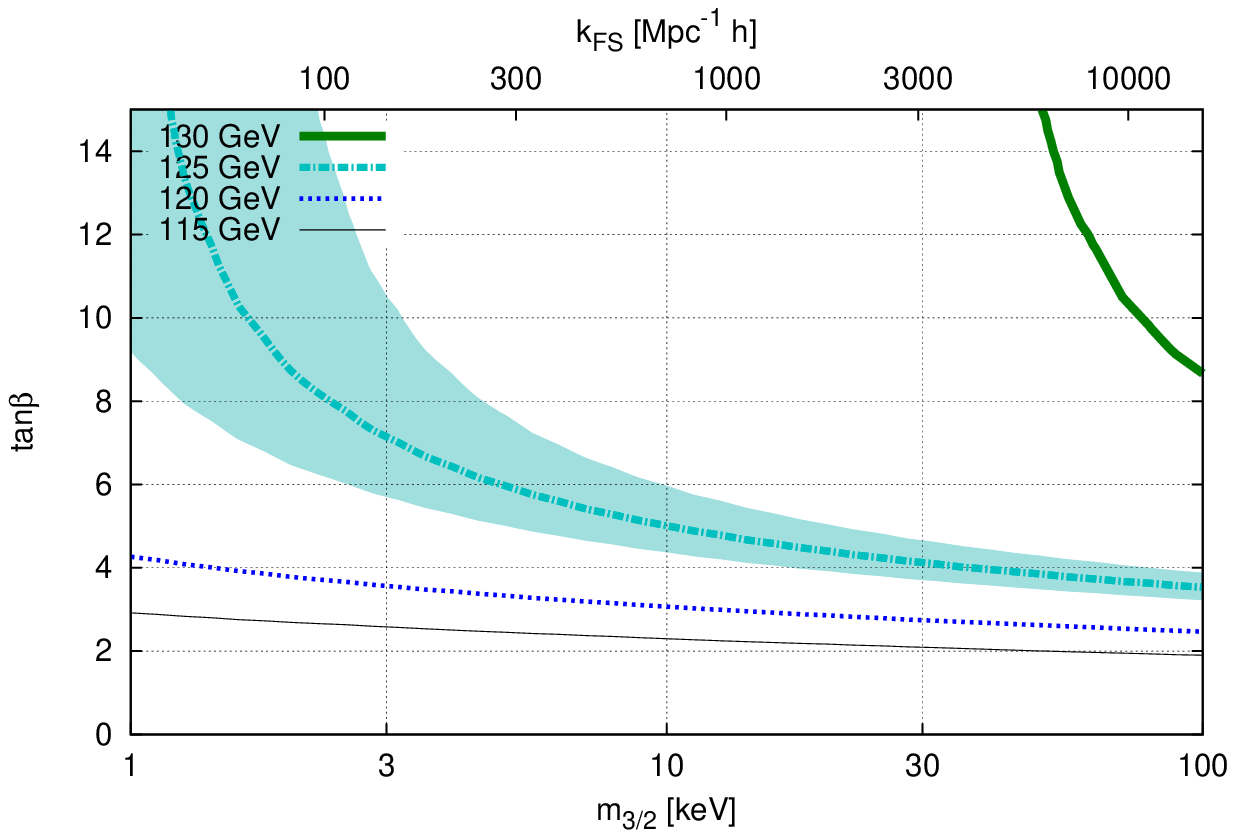}
  \end{center}
  \begin{center}
   \includegraphics{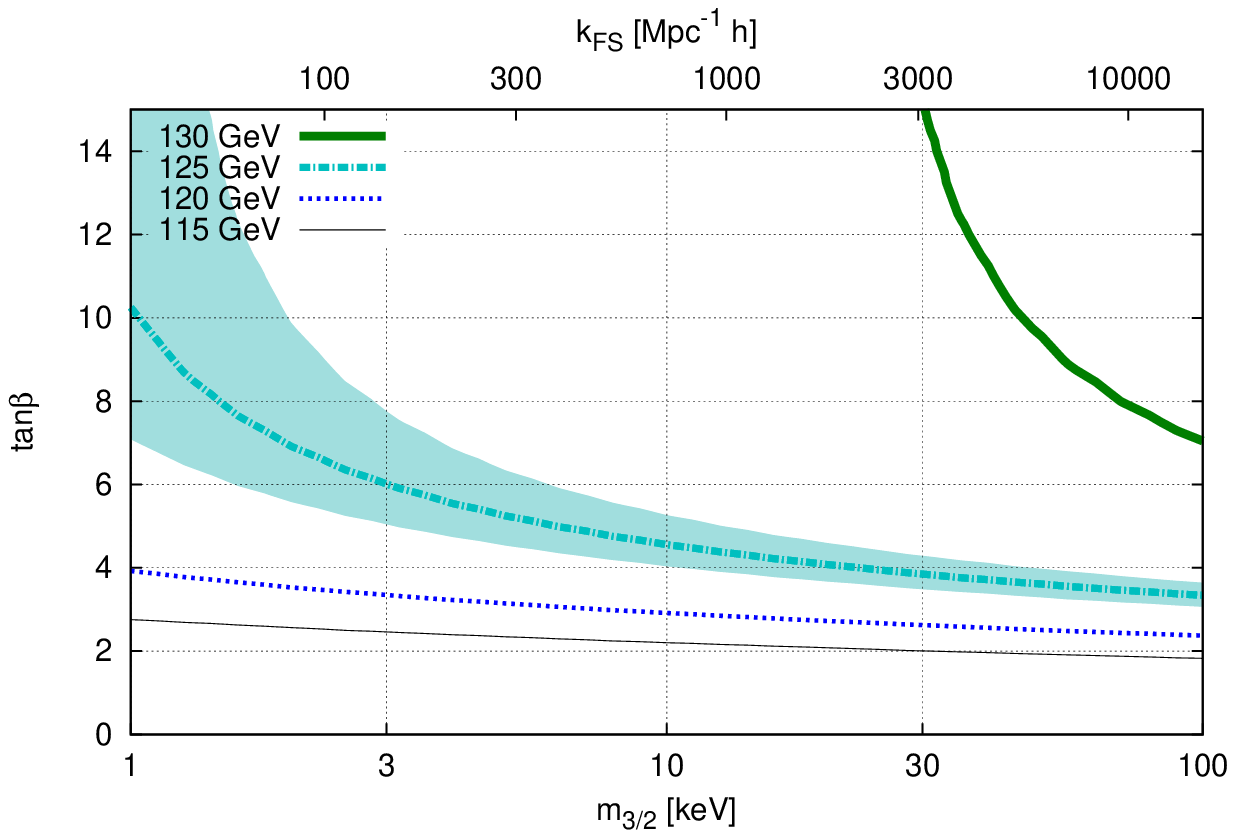}
  \end{center}
 \caption{The SM-like Higgs boson mass in one pair of messenger model (top)
 and two pairs of messenger model (bottom).
 In the filled region, the Higgs mass satisfies $124~\GEV \leq m_h \leq 126~\GEV$.
 We take $k_\ell F/m_\ell^2 = 0.95$ and $\langle Z\rangle / m_\ell = 1$.}
 \label{fig:higgsmass}
\end{figure}

In Fig.\,\ref{fig:higgsmass},  we show the Higgs boson mass obtained 
by solving the renormalization group equations by matching the MSSM
and the SM with gauginos with the boundary condition, 
\begin{eqnarray}
\label{eq:SUSY}
  \lambda = \frac{1}{4} \left(\frac{3}{5}g_1^2+ g_2^2 \right) \cos^22\beta\ ,
\end{eqnarray}
at the stop mass scale.%
\footnote{
In this section, we have assumed the MSSM to obtain the lightest Higgs boson mass.
As we will discuss in the next section, however, it is rather difficult to achieve correct 
electroweak symmetry breaking in the MSSM with gauge mediated supersymmetry breaking
for squark masses in $O(10-100)$\,TeV.
Correct electroweak symmetry breaking is, on the other hand, realized with 
an introduction of a singlet coupling to the Higgs doublet.
Such interaction does not alter the following discussion based 
on the MSSM, since the required coupling between the singlet and the Higgs doublets 
is very small (see next section).
}
We also match the renormalization group equations of the models with and 
without gauginos at the gaugino mass scales.
The threshold corrections at the heavy scalar scale 
and at the weak scale are also taken into account.
in accordance with Refs.\,\cite{Bernal:2007uv,Giudice:2011cg}.%
\footnote{In our analysis, we also assumed that Higgsinos are as heavy 
as the sfermions.
The effects of such a heavy Higgsinos to the quartic coupling constant of the 
Higgs doublet are discussed in Ref.\,\cite{Ibe:2011aa}.
}

In our calculation, we take $m_Z$ scale gauge coupling constants as
$\a(m_Z)^{-1} = 128.944$
\,\cite{Hagiwara:2011af} and
$\a_s(m_Z) = 0.1184$
\,\cite{Bethke:2009jm}.
We use a formula given in Ref.\,\cite{Chankowski:1994ua} for 
the Weinberg angle $\sin^2 \theta_W(m_Z)$.
The top Yukawa coupling constant $y_t$ is calculated by using top quark pole mass 
$m_t = 173.2~\GEV$
\,\cite{Lancaster:2011wr} and 
weak scale threshold corrections given in Ref.\,\cite{Chetyrkin:1999qi}.

The figure shows that the model leads to the lightest Higgs boson mass 
around $125$\,GeV which is suggested by recent ATLAS and CMS results
for $m_{3/2}=1-100$\,keV for moderate values of the mixing angle 
of the two Higgs doublets, $\tan\b$.
We also show the free-streaming wavenumber for a given gravitino 
mass.
As a result, we find that the both the Higgs boson mass around $125$\,GeV
and the free-streaming wavenumber $k_{FS}\simeq 100-300$\,Mpc$^{-1} h$
can be achieved simultaneously for $m_{3/2}={\cal O}(1)$\,keV.

\section{The origins of $\mu$/$B\mu$}\label{sec:ew}
Finally, let us discuss the origins of the supersymmetric and 
the supersymmetry breaking Higgs mixing 
parameters,  the so-called $\mu$ and $B$ parameters.
As we have seen in section\,\ref{sec:model}, the scalar boson masses
are around $10$\,TeV.
In such cases, the $\mu$ parameter is also required to be of $O(10)$\,TeV
for successful electroweak symmetry breaking.

The simplest way to provide the $\mu$ parameter is to assume
that $\mu$ parameter is given by hand, i.e.,
\begin{eqnarray}
 W = \mu H_u H_d \ .
\end{eqnarray}
In this case, the $B$ parameter is mainly generated by the renormalization group effects,
 \begin{eqnarray}
 \frac{d B}{d \ln \mu_R}  = 
 \frac{1}{16\pi^2} 
\left[
6 a_t y_t
+ 6 a_b y_b
+ 6 g_2^2 M_2 
+ \frac{6}{5}g_1^2 M_1
\right] 
\simeq  \frac{1}{16\pi^2} 
\left[
 6 g_2^2 M_2 
+ \frac{6}{5}g_1^2 M_1
\right] \ .
\end{eqnarray}
Here, $\m_R$ denotes the renormalization scale,
 $y_{t,b}$ the top and the bottom Yukawa coupling constants,
and $a_{t,b}$ the $A$-terms of the top and the bottom squarks.
In the final expression, we have neglected the $A$-term contributions 
to the beta function of $B$, since $A$-terms are generated at the two-loop order
in the gauge mediation models.
As a result, the radiatively generated $B$-term in this model is roughly given by 
\begin{eqnarray}
 B \sim \frac{1}{16\pi^2} 
\left[
 6 g_2^2 M_2 
+ \frac{6}{5}g_1^2 M_1
\right]  \log\frac{M_{\rm mess}}{m_{\rm squark}}
= O(0.1) \times M_2 \ ,
\end{eqnarray}
where we have used the messenger scale, $M_{\rm mess } = O(10^7)$\,GeV, 
and $m_{\rm squark} = O(10)$\,TeV.
Unfortunately, however, the above generated $B$ term is too small 
to be viable since it leads a too  large  Higgs mixing angle $\tan\b$ via,
\begin{eqnarray}
\label{eq:sin2b}
\sin2\b = \frac{2 B \m}{m_{H_u}^2 + m_{H_d}^2 + 2 \m^2} \simeq O(10^{-2})\ .
\end{eqnarray}
Such a large $\tan\b$ causes a Landau pole problem of the top Yukawa coupling 
below the GUT scale.
Therefore, we need an alternative mechanism 
for the generations of the $\mu$ term and $B$ term
for a successful model.

\begin{table}
\label{tab:spectrum}
\begin{center}
\begin{tabular}{| l | cl || l | cl |} 
\hline\hline
 ${h_1^0}$        & 125&GeV      & ${\tilde{l}_L}$  & 17&TeV   \\
 ${h_2^0}$        & 7.0&TeV      & ${\tilde{e}_R}$  & 9.6&TeV   \\
 ${h_3^0}$        & 22&TeV     & ${\tilde{\tau}_1}$  & 9.6&TeV\\
 ${A_1^0}$        & 12&TeV     & ${\tilde{\tau}_2}$  & 17&TeV\\
 ${A_2^0}$        & 22&TeV      & ${\tilde{\nu}}$  & 17&TeV   \\
 ${H^\pm}$        & 22&TeV      & ${\tilde{\nu}_\tau}$  & 17&TeV   \\
 ${\chi_1^\pm}$   & 2.7&TeV     & ${\tilde{u}_L}$  & 39&TeV  \\
 ${\chi_2^\pm}$   & 14&TeV      & ${\tilde{u}_R}$  & 36&TeV   \\
 ${\chi_1^0}$     & 0.97&TeV    & ${\tilde{t}_1}$  & 32&TeV  \\
 ${\chi_2^0}$     & 2.7&TeV      & ${\tilde{t}_2}$  & 37&TeV  \\
 ${\chi_3^0}$     & 7.1&TeV     & ${\tilde{d}_L}$  & 39&TeV    \\
 ${\chi_4^0}$     & 14&TeV      & ${\tilde{d}_R}$  & 36&TeV   \\
 ${\chi_5^0}$     & 14&TeV      & ${\tilde{b}_1}$  & 36&TeV   \\
 ${\tilde{g}}$    & 1.9&TeV      & ${\tilde{b}_2}$  & 37&TeV \\
\hline\hline
\end{tabular}
\hspace{1cm}
\begin{tabular}{|l|l|} 
\hline\hline
 $\tan\b$  & 5  \\
 $\l_H(m_{\rm stop})$  & $3\times 10^{-2}$  \\
 $\kappa(m_{\rm stop})$  & $-7.4\times 10^{-3}$ \\
 $\xi_F(M_{\rm mess})$  & $4\times 10^9\,{\rm GeV}^2$ \\
  ${m_S^2(M_{\rm mess})}$  & $5.0\times 10^7{\rm GeV}^2$ \\
\hline\hline
\end{tabular}
\end{center}
\caption{A sample mass spectrum in the extended NMSSM 
is shown for $m_{3/2}\simeq 7$\,keV. 
We introduce two pairs of messengers,
and take $k_\ell F/m_\ell^2 = 0.95$ and $\langle Z \rangle / m_\ell = 1$.
In our analysis, we have set trilinear soft supersymmetry breaking terms 
are vanishing at the messenger scale, while the linear soft supersymmetry breaking 
term to be generated by the supergravity effects, i.e. $V_{\rm soft} = -2 m_{3/2}\xi_F S + h.c.$.
}
\end{table}

The next to the MSSM (NMSSM), on the other hand, provides 
 the effective $\m$ and $B$ parameters successfully.
The superpotential of the Higgs sector in the (extended) NMSSM is given by
  \begin{eqnarray}
  \label{eq:NMSSM}
 W = \l_H S H_u H_d +\xi_F S + \frac{\k}{3} S^3\ ,
\end{eqnarray}
where $S$ is a gauge singlet Higgs field\,\cite{Morrissey:2008gm}. 
We introduced 
dimensionless coupling constant $\l_H$ and $\k$, and also 
a dimensionful parameter  $\xi_F$.
It should be noted that this superpotential is symmetric under a discrete $R$-symmetry,
$Z_{4R}$, with the charge assignments $Q(S) = 2$, which can be consistent
with R-symmetry of the messenger sector in the previous section.
In the extended NMSSM, the effective $\mu$ and
$B$ parameters are provided by
the vacuum expectation values of $S$ and its $F$-component,
\begin{eqnarray}
  \mu_{\rm eff}  &=& \l_H \vev{S} \ , \\
  B_{\rm eff}  &=& \vev{F_S}/{\vev S} = \k \vev{S}   + \xi_F/\vev{S}\ . 
\end{eqnarray}
Here, the vacuum expectation value of $S$ is obtained by the scalar potential of $S$,
\begin{eqnarray}
 V =  m_{S}^2 |S|^2 + |\xi_F + \k S^2|^2 \ ,
\end{eqnarray}
with the soft mass parameter $m_S^2$ leading to%
\footnote{Here, we have assumed that all the parameters are real valued.},
\begin{eqnarray}
  \vev S \sim \frac{\sqrt{-m_S^2-2 \k \xi_F}}{\sqrt 2 \k} \ .
\end{eqnarray}

In table\,\ref{tab:spectrum}, we show a sample spectrum of the model
for $m_{3/2}\simeq 7\,$keV.
To obtain the spectrum, we used {\it NMSSMTool}\,\cite{NMSSMtools}  modified for 
our purpose.
In our analysis, we  have further assumed that $m_S^2$ is generated 
by the interactions between $S$ and heavier fields such as
extra quark multiplets\,\cite{deGouvea:1997cx}.
To parametrize such effects, we simply introduced $m_S^2$ at the messenger
scale.
It should be noted that 
$m_S^2$ is expected to be one-loop suppressed compared with
squark mass squared if $m_S^2$ is generated via the interactions
to the extra quark multiplets.
The table shows that the electroweak symmetry breaking 
is successfully obtained in the extended NMSSM even for such 
a suppressed $m_S^2$, i.e. $m_{S}^2/m_{\rm squark}^2 = O(10^{-2})$.

Before closing this section, let us comment on the effect of 
 the interaction between $S$ to the Higgs doublets in Eq.\,(\ref{eq:NMSSM})
on the lightest Higgs boson mass,
\begin{eqnarray}
 {\mit \D}m_{\rm higgs}^2 = \l_H^2 v^2 \sin^2 2\b \ .
\end{eqnarray}
This contribution is, however, highly suppressed for $\tan\b = O(10)$ and $\l_H \ll O(1)$.
Therefore, the additional contribution does not change 
the correlation between the Higgs boson mass 
and the free-streaming length of the gravitino in the previous section 
is not changed even if we assume the extended NMSSM.

\section{Conclusions}
In this Letter, we discussed the SM-like Higgs boson mass in the supersymmetric standard model
in the R-invariant direct gauge mediation model with dark matter consisting 
of the gravitino with mass in the ${\cal O}(1)$\,keV range.
With such a gravitino mass, 
the model predicts rather heavy sfermion masses
in ${\cal O}(10-100)\,\TEV$,
which leads to a rather heavy lightest Higgs boson mass.
As a result, we showed that the Higgs boson mass around 125\,GeV
can be easily achieved, which is suggested by the ATLAS and CMS experiments.
Interestingly, gravitino dark matter with this mass range
is an attractive candidate for warm dark matter to surmount
some difficulties of the cold dark matter scenario
at the small scale structure.

One of the distinctive features of the R-symmetric direct mediation model is that 
the model predicts a peculiar gaugino mass spectrum which 
violates the so-called GUT relation,
in spite of the GUT invariant boundary condition.
Thus, the measurement of the gaugino mass spectrum at the collider experiments
provide a consistency test of the model.
The null observation of the sfermions is also an important prediction
of the present model.
The warmness of  dark matter will be more disclosed 
in future by the progress of the $N$-body simulation
of the structure formation as well as the observation of the structure of the universe.

\section*{Acknowledgements}

We would like to thank T.T. Yanagida for suggesting the subject 
and for useful discussions at the early stage of the project.
We would also like to thank A. Kamada for useful comments and discussions
on the properties of warm dark matter.
The work of MI is supported by Grant-in-Aid for Scientific research from the Ministry of Education, Science, Sports, and Culture (MEXT), Japan, No. 24740151.
The work of RS is supported in part by JSPS Research
Fellowships for Young Scientists.
This work  is also supported by the World Premier 
International Research Center Initiative (WPI Initiative), MEXT, Japan.


\begin{thebibliography}{99}


\bibitem{Viel:2005qj} 
  M.~Viel, J.~Lesgourgues, M.~G.~Haehnelt, S.~Matarrese and A.~Riotto,
  Phys.\ Rev.\ D {\bf 71}, 063534 (2005)
  [astro-ph/0501562].


\bibitem{WDM}
For recent developments on the effects on the galaxy formation in the warm dark matter scenario,
see, for example, 
  H.~J.~de Vega, P.~Salucci and N.~G.~Sanchez,
  New Astron.\  {\bf 17}, 653 (2012)
  [arXiv:1004.1908 [astro-ph.CO]];
  K.~Markovic, S.~Bridle, A.~Slosar and J.~Weller,
  JCAP {\bf 1101}, 022 (2011)
  [arXiv:1009.0218 [astro-ph.CO]];
  D.~Boyanovsky,
  Phys.\ Rev.\ D {\bf 83}, 103504 (2011)
  [arXiv:1011.2217 [astro-ph.CO]];
  A.~Kamada and N.~Yoshida, in preparation,
  and references therein. 


\bibitem{Ibe:2010ym} 
  M.~Ibe, R.~Sato, T.~T.~Yanagida and K.~Yonekura,
  JHEP {\bf 1104}, 077 (2011)
  [arXiv:1012.5466 [hep-ph]].

\bibitem{Izawa:1997gs} 
  K.~I.~Izawa, Y.~Nomura, K.~Tobe and T.~Yanagida,
  Phys.\ Rev.\ D {\bf 56}, 2886 (1997)
  [hep-ph/9705228].

\bibitem{Nomura:1997uu} 
  Y.~Nomura and K.~Tobe,
  Phys.\ Rev.\ D {\bf 58}, 055002 (1998)
  [hep-ph/9708377].
\bibitem{Fujii:2003iw} 
  M.~Fujii, M.~Ibe and T.~Yanagida,
  Phys.\ Rev.\ D {\bf 69}, 015006 (2004)
  [hep-ph/0309064].

\bibitem{Hasenkamp:2010if} 
  J.~Hasenkamp and J.~Kersten,
  Phys.\ Rev.\ D {\bf 82}, 115029 (2010)
  [arXiv:1008.1740 [hep-ph]].

\bibitem{Ibe:2005xc} 
  M.~Ibe, K.~Tobe and T.~Yanagida,
  Phys.\ Lett.\ B {\bf 615}, 120 (2005)
  [hep-ph/0503098].
\bibitem{Shih:2007av} 
  D.~Shih,
  JHEP {\bf 0802}, 091 (2008)
  [hep-th/0703196].
\bibitem{Komargodski:2009jf} 
  Z.~Komargodski and D.~Shih,
  JHEP {\bf 0904}, 093 (2009)
  [arXiv:0902.0030 [hep-th]].
\bibitem{Sato:2009dk} 
  R.~Sato and K.~Yonekura,
  JHEP {\bf 1003}, 017 (2010)
  [arXiv:0912.2802 [hep-ph]].

\bibitem{Okada:1990gg} 
  Y.~Okada, M.~Yamaguchi and T.~Yanagida,
  Phys.\ Lett.\ B {\bf 262}, 54 (1991).


\bibitem{:2012si} 
  G.~Aad {\it et al.}  [ATLAS Collaboration],
  arXiv:1202.1408 [hep-ex].
  
\bibitem{Chatrchyan:2012tx} 
  S.~Chatrchyan {\it et al.}  [CMS Collaboration],
  arXiv:1202.1488 [hep-ex].
  
  \bibitem{SUSY}
 ATLAS report,  ATLAS-CONF-2012-033
 
\bibitem{Komatsu:2010fb} 
  E.~Komatsu {\it et al.}  [WMAP Collaboration],
  Astrophys.\ J.\ Suppl.\  {\bf 192}, 18 (2011)
  [arXiv:1001.4538 [astro-ph.CO]].


 
\bibitem{Izawa:1996pk} 
  K.~-I.~Izawa and T.~Yanagida,
  Prog.\ Theor.\ Phys.\  {\bf 95}, 829 (1996)
  [hep-th/9602180].
\bibitem{Intriligator:1996pu} 
  K.~A.~Intriligator and S.~D.~Thomas,
  Nucl.\ Phys.\ B {\bf 473}, 121 (1996)
  [hep-th/9603158].





\bibitem{Bond:1980ha} 
  J.~R.~Bond, G.~Efstathiou and J.~Silk,
  Phys.\ Rev.\ Lett.\  {\bf 45}, 1980 (1980).
\bibitem{Kamada}  
Private communication with A.~Kamada.

\bibitem{Klypin:1999uc} 
  A.~A.~Klypin, A.~V.~Kravtsov, O.~Valenzuela and F.~Prada,
  Astrophys.\ J.\  {\bf 522}, 82 (1999)
  [astro-ph/9901240];
  B.~Moore, S.~Ghigna, F.~Governato, G.~Lake, T.~R.~Quinn, J.~Stadel and P.~Tozzi,
  Astrophys.\ J.\  {\bf 524}, L19 (1999)
  [astro-ph/9907411].
\bibitem{Strigari:2007ma} 
  L.~E.~Strigari, J.~S.~Bullock, M.~Kaplinghat, J.~Diemand, M.~Kuhlen and P.~Madau,
  Astrophys.\ J.\  {\bf 669}, 676 (2007)
  [arXiv:0704.1817 [astro-ph]].
\bibitem{deVega:2009ku} 
  H.~J.~de Vega and N.~G.~Sanchez,
  Mon.\ Not.\ Roy.\ Astron.\ Soc.\  {\bf 404}, 885 (2010)
  [arXiv:0901.0922 [astro-ph.CO]].
  \bibitem{WDM2}
See  New Astron.\  {\bf 17}, 653 (2012) in Ref.\,\cite{WDM}.
\bibitem{Dalal:2002su} 
  N.~Dalal and C.~S.~Kochanek,
  [astro-ph/0202290].
\bibitem{Loeb:2003ya} 
  A.~Loeb and M.~Zaldarriaga,
  Phys.\ Rev.\ Lett.\  {\bf 92}, 211301 (2004)
  [astro-ph/0312134].



\bibitem{Bernal:2007uv} 
  N.~Bernal, A.~Djouadi and P.~Slavich,
  JHEP {\bf 0707}, 016 (2007)
  [arXiv:0705.1496 [hep-ph]].
\bibitem{Giudice:2011cg} 
  G.~F.~Giudice and A.~Strumia,
  Nucl.\ Phys.\ B {\bf 858}, 63 (2012)
  [arXiv:1108.6077 [hep-ph]].
\bibitem{Ibe:2011aa} 
  M.~Ibe and T.~T.~Yanagida,
  Phys.\ Lett.\ B {\bf 709}, 374 (2012)
  [arXiv:1112.2462 [hep-ph]];
  M.~Ibe, S.~Matsumoto and T.~T.~Yanagida,
  arXiv:1202.2253 [hep-ph].




\bibitem{Hagiwara:2011af} 
  K.~Hagiwara, R.~Liao, A.~D.~Martin, D.~Nomura and T.~Teubner,
  J.\ Phys.\ G G {\bf 38}, 085003 (2011)
  [arXiv:1105.3149 [hep-ph]].
  
\bibitem{Bethke:2009jm} 
  S.~Bethke,
  Eur.\ Phys.\ J.\ C {\bf 64}, 689 (2009)
  [arXiv:0908.1135 [hep-ph]].

\bibitem{Chankowski:1994ua} 
  P.~H.~Chankowski, Z.~Pluciennik and S.~Pokorski,
  Nucl.\ Phys.\ B {\bf 439}, 23 (1995)
  [hep-ph/9411233].
  
\bibitem{Lancaster:2011wr} 
  [Tevatron Electroweak Working Group and for the CDF and D0 Collaborations],
  arXiv:1107.5255 [hep-ex].
  
\bibitem{Chetyrkin:1999qi} 
  K.~G.~Chetyrkin and M.~Steinhauser,
  Nucl.\ Phys.\ B {\bf 573}, 617 (2000)
  [hep-ph/9911434].
  
\bibitem{Morrissey:2008gm} 
For more detailed discussion on the NMSSM in the gauge mediation models,
see, for example,
  D.~E.~Morrissey and A.~Pierce,
  Phys.\ Rev.\ D {\bf 78}, 075029 (2008)
  [arXiv:0807.2259 [hep-ph]];
  U.~Ellwanger, C.~-C.~Jean-Louis and A.~M.~Teixeira,
  JHEP {\bf 0805}, 044 (2008)
  [arXiv:0803.2962 [hep-ph]],
  and references therein.
  
\bibitem{NMSSMtools}  
  U.~Ellwanger, J.~F.~Gunion and C.~Hugonie,
  JHEP {\bf 0502}, 066 (2005)
  [hep-ph/0406215];
  U.~Ellwanger and C.~Hugonie,
  Comput.\ Phys.\ Commun.\  {\bf 175}, 290 (2006)
  [hep-ph/0508022].
  see also http://www.th.u-psud.fr/NMHDECAY/nmssmtools.html
  
  
\bibitem{deGouvea:1997cx} 
  A.~de Gouvea, A.~Friedland and H.~Murayama,
  Phys.\ Rev.\ D {\bf 57}, 5676 (1998)
  [hep-ph/9711264];
  M.~Ibe, R.~Kitano and H.~Murayama,
  Phys.\ Rev.\ D {\bf 71}, 075003 (2005)
  [hep-ph/0412200].
\end{thebibliography}
\end{document}